\documentclass[11pt,draft,onecolumn,twoside]{IEEEtran}
\usepackage{cite}
\usepackage{amsmath,amssymb,amsfonts,amsthm,mathrsfs}
\usepackage{multirow}
\usepackage{bbm}
\usepackage{algorithmic}
\usepackage{caption}
\usepackage{graphicx}
\usepackage{tikz} 
\usetikzlibrary{automata, positioning, arrows}
\usepackage{textcomp}
\usepackage{xcolor}
\usepackage{manfnt}
\usepackage{mathtools}
\usepackage{fixmath}
\usepackage{pgf}
\usepackage{xcolor}
\usetikzlibrary{positioning,arrows}
\usepackage{balance} 

\def\BibTeX{{\rm B\kern-.05em{\sc i\kern-.025em b}\kern-.08em
    T\kern-.1667em\lower.7ex\hbox{E}\kern-.125emX}}
    
\newcommand{\N}{\mathbb{N}}

\newcommand{\R}{\mathbb{R}}
\newcommand{\Q}{\mathbb{Q}}

\newcommand{\sA}{\mathcal{A}}
\newcommand{\sC}{\mathcal{C}}
\newcommand{\sM}{\mathcal{M}}

\newcommand{\sT}{\mathcal{T}}

\newcommand{\sX}{\mathcal{X}}
\newcommand{\sY}{\mathcal{Y}}

\newcommand{\fT}{\mathfrak{T}}

\theoremstyle{plain}
\newtheorem{thm}{Theorem}
\newtheorem{lem}{Lemma}

\newtheorem{cor}{Corollary}

\theoremstyle{defn}
\newtheorem{defn}{Definition}

\theoremstyle{rem}
\newtheorem{rem}{Remark}

\tikzstyle{block} = [draw, rectangle, 
  minimum height=3em, minimum width=4em]

\newcommand{\tcr}[1]{\textcolor{red}{#1}}

\newcommand{\tcb}[1]{\textcolor{blue}{#1}}
\newcommand{\tco}[1]{\textcolor{orange}{#1}}

\renewcommand{\tcr}[1]{#1}
\renewcommand{\tco}[1]{#1}
\renewcommand{\tcb}[1]{#1}

\begin{document}

\title{Algorithmic Computability of the Capacity of Gaussian Channels with Colored Noise}
\author{
Holger Boche, Andrea Grigorescu, Rafael F. Schaefer, and H. Vincent Poor
}
\maketitle

\begin{abstract}
Designing capacity-achieving coding schemes for the band-limited additive colored Gaussian noise (ACGN) channel has been and is still a challenge. In this paper, the capacity of the band-limited ACGN channel is 
studied from a fundamental algorithmic point of view by addressing the question of whether or not the capacity can be algorithmically computed. To this aim, the concept of Turing machines is used, which provides fundamental performance limits of digital computers. 
t is shown that there are band-limited ACGN channels having computable continuous spectral densities whose capacity are non-computable numbers. Moreover, it is demonstrated that for those channels, it is impossible to find computable sequences of asymptotically sharp upper bounds for their capacities.
\end{abstract}

\footnotetext[1]{This work of H. Boche was supported in part by the German Federal Ministry of Education and Research (BMBF) within the national initiative on 6G Communication Systems through the research hub 6G-life under Grant 16KISK002, within the national initiative on Post Shannon Communication
(NewCom) under Grant 16KIS1003K, and the project Hardware Platforms and Computing Models for Neuromorphic Computing (NeuroCM) under Grant 16ME0442. It has further received funding by the German Research Foundation (DFG) within Germany’s Excellence Strategy EXC-2092 – 390781972.
 This work of R. F. Schaefer was supported in part by the BMBF within NewCom under Grant 16KIS1004 and 6G-life under Grant 16KISK001K as well as in part by the German Research Foundation (DFG) under Grant SCHA 1944/11-1.}
 
\footnotetext[2]{H. Boche is with the Chair of Theoretical Information Technology, Technical University of Munich, Arcisstr. 21, 80333 M\"unchen, Germany. (E-mail:boche@tum.de). He is also with the BMBF Research Hub 6G-life and with the Excellence Cluster Cyber Security in the Age of Large-Scale Adversaries (CASA), Ruhr University Bochum.
A. Grigorescu is with the Chair of Theoretical Information Technology, Technical University of Munich, Arcisstr. 21, 80333 M\"unchen, Germany. (E-mail:andrea.grigorescu@tum.de).
R. F. Schaefer is with the Chair of Information Theory and Machine Learning, Technische Universit\"at Dresden, the BMBF Research Hub 6G-life, the Cluster of Excellence ``Centre for Tactile Internet with Human-in-the-Loop (CeTI)'', and the 5G Lab Germany, Technical University of Dresden, 01069 Dresden, Germany (E-mail: rafael.schaefer@tu-dresden.de).
H.V. Poor is with the Department of Electrical and Computer Engineering, Princeton University (E-mail:poor@princeton.edu)}

\section{Introduction}
While the 5th generation (5G) of mobile networks enable the industrial application of the Internet of Things (IoT), the technological advances of the wireless network infrastructure aimed for 6G will provide the basis to massively expand the use of IoT and Tactile Internet for consumers, creating innovative approaches that improve people's quality of life. This will not only substantially increase the amount of data traffic over wireless networks, but also means that physical and virtual objects will be controlled over the network. Hence, the 6G infrastructure will have to provide reliable, scalable and secure communication with substantially higher throughput than 5G does, and with low latency. The advances in wireless network infrastructure for 6G will have to enable sensing and coordination of control functions; see \cite{fettweis20216gtactile, fettweis20226gtrust}. On the other hand, as any technological advances that enhance sensory capabilities can potentially be exploited by malicious actors. 6G networks will thus require architectural solutions guaranteeing not only security, but also legal and social requirements, such as the General Data Protection Regulation (GDPR). Therefore, it is crucial to build a native, trustworthy architecture for 6G networks.

The trustworthiness framework for 6G consists of privacy, security, integrity, resilience, reliability, availability, accountability, authenticity, and device independence. In this paper, we will approach the reliability and integrity aspects of trustworthiness. In particular, the reliability aspect of communication networks has been studied intensively in information theory.

The goal of information theory is to provide a mathematical framework to model communication scenarios and to quantify their properties. The information-theoretic framework allows for the derivation of benchmarks that determine the reliable transmission rates for a communication channel while considering its characteristics, noise, and power constraints. These benchmarks enable the establishment of reliable information transfer rates for the channel under consideration and can be used in the design of communication systems to find optimal coding strategies operating at high transmission rates with minimal transmission errors.

A crucial objective of information theory is to formulate rate benchmarks as optimization problems using elementary functions, ideally expressed in closed form, which can provide precise and concise representations of the solutions, allowing for ease of computation and analysis.

In the field of information theory, researchers aim to find simple capacity formulas, which are then evaluated numerically to compute the actual capacity of the system. Numerical evaluation is a crucial step in this process. Computing capacity results are highly significant in practice, as they serve as benchmarks for the development of real-world communication systems.

Computing rate benchmarks using digital computers, such as the capacities of channels or channel reliability functions, has long been an area of interest. Some performance functions are implicitly assumed to be computable, particularly those that involve entropic quantities such as capacity expressions. In 1967, methods for constructing both upper and lower bounds for channel reliability functions were introduced in \cite{shannon1967lower}. These techniques were specifically designed to enable the computation of these bounds using digital computers.

 In 1972, an algorithm to compute the capacity of arbitrary discrete memoryless channels (DMCs) was independently presented in \cite{arimoto1972algorithm} and \cite{blahut1972computation}. In \cite{blahut1972computation}, an analogous algorithm was proposed to compute the rate distortion trade-off of lossy source compression. In general, the capacity is usually given in terms of mutual information expressions. Note that even for the  binary symmetric channel (BSC) with a rational crossover probability, i.e., $\epsilon\in(0,\frac{1}{2})\cap\Q$, the capacity is a transcendental number \cite{boche2022algorithmic}. Hence, a precise calculation is not possible, since the calculation has to stop after a finite number of computation steps. Only a suitable approximation of it can be calculated.

The algorithmic computability properties of capacity has been studied for various channels, including finite-state channels (FSCs) \cite{boche2020shannon}, FSCs with feedback \cite{grigorescu2022capacity}, compound channels \cite{boche2020communication}, and correlation-assisted DMCs \cite{boche2019identification}. For all of these channels, it has been demonstrated that the capacities are not generally computable functions, due to their complicated descriptions. This prompts an interesting question:\emph{ What is the simplest communication channel for which such a numerical computation of the capacity is not possible?} We provide an answer to this query by showing that there are band-limited additive colored Gaussian noise (ACGN) channels, which is a standard communication channel with a very simple structure, that do not have a computable capacity.

The band-limited ACGN channel is a very important model for wireless communication, as it can be used to model commonly encountered channels such as the frequency selective fading channel. The band-limited Gaussian channel, introduced in \cite{shannon1949communication, shannon1949communicationnoise}, is a continuous-time channel. In \cite{shannon1949communicationnoise}, two noise models are introduced: white and colored Gaussian noise. Colored Gaussian noise is Gaussian distributed and has a power spectral density (p.s.d.) that varies with frequency while the spectral density of white Gaussian noise is a constant over the frequency. The capacity and error performance of codes for the band-limited ACGN channel were carefully studied in \cite{wyner1966capacity, ash1963capacity, shannon1959probability}. A detailed description of the band-limited ACGN channel and its results can be found in \cite{ash2012information,cover1999elements,ihara1993information,tse2005fundamentals,pinsker1964information}. In \cite{gallager1968information}, Gallager showed that the capacity-achieving p.s.d. of the linear ACGN channel can be determined using the water pouring technique. In \cite{forney1998modulation}, the authors provide an overview of techniques desired to construct capacity-achieving codes for the ACGN channel. 

In addition, the capacity function of the  band-limited ACGN channel plays a fundamental role in multi-user information theory, e.g., the broadcast channel with intersymbol interference (ISI) \cite{goldsmith2001capacity}, Gaussian broadcast channel \cite{hughes1975capacity,poltyrev1979capacity,el1980capacity}, discrete time Gaussian channel with ISI \cite{hirt1988capacity}, multiple access channel (MAC) with memory \cite{verdu1989multiple}, Gaussian MAC with ISI \cite{cheng1993gaussian}, and others.

Computing the capacity of the band-limited ACGN channel is a very important task for practical systems and, especially, for 6G. The capacity serves as a benchmark for designing and optimizing systems to achieve optimal message transmission. This will enable the design of codes that fulfill anticipated reliability and efficiency requirements of 6G systems, while also optimizing the use of communication resources. For instance, resource allocation will most likely take place at the base stations of 6G systems. At the present and in the foreseeable future, only digital hardware can be used for this task. 

To compute the capacity of band-limited ACGN channels on digital computers, a program must be able to compute a description of the capacity from its parameters. However, the capacity of band-limited ACGN channels is generally not computable. Thus, making it impossible for any algorithm to fulfill the integrity requirement and ensure the trustworthiness of the communication system through digital hardware.

 To address algorithmic computability, we use the concept of a \emph{Turing machine} \cite{turing1936computable,turing1938computable,weihrauch2000computable}, which is a mathematical model of an abstract machine that manipulates symbols on a strip of tape according to certain given rules. Any algorithm can be translated into a sequence of steps that can be executed by a Turing machine and therefore provides a simple and very powerful model of computation. Turing machines have no limitations on computational complexity, computing capacity or storage, and execute programs completely error-free. Accordingly, they provide fundamental performance limits for today's digital computers. Turing machines account for all those problems and tasks that are algorithmically computable on a classical (i.e., non-quantum) machine. They are equivalent to the von Neumann-architecture without hardware limitations and the theory of recursive functions \cite{godel1930vollstandigkeit,godel1934undecidable,kleene1952introduction,minsky1961recursive,avigad2014computability}. 
 
This paper addresses the question of whether the capacity of the band-limited ACGN channel can be computationally determined. We demonstrate that there exist band-limited ACGN channels for which it is impossible to find an algorithm that approximates its capacity within any desired margin of error. Additionally, we prove that it is impossible to find algorithmically computable upper bounds on the capacity of band-limited ACGN channels. Finally, we show that relaxing the power constraint does not improve the computational behavior of the capacity.

The remainder of the paper is organized as follows: In Section \ref{sec:Gauss_Channel}, we introduce the band-limited ACGN channel and present its capacity results. In Section \ref{sec:computability_numbers}, we introduce the computability framework. In Section \ref{sec:computability_Gauss}, we show the existence of band-limited ACGN channels whose capacities yields a non-computable numbers. Finally, we provide our conclusions in Section \ref{sec:conclusion}.

\section{Continuous Gaussian Channels}
\label{sec:Gauss_Channel}
In this section we consider a communication scenario where both the input and output of the channel are amplitude- and time-continuous. Amplitude-continuous means that the signal alphabets are uncountably infinite, and by time-continuous we allow the transmission to be continuous over time.
The time continuous additive Gaussian channel is represented by the following formula:
\begin{equation*}
	y(t)=x(t)+ z(t), 
\end{equation*}
where $x(t), y(t)$, and $z(t)$ are the channel input, channel output and noise at time instant $t\in\sT\subset\R$ and they take values in $\R$.  The noise $z(t)$ is zero-mean Gaussian distributed. The Fourier \tco{transforms} of the input signals and noise are represented by
\begin{align*}
X(f)&=\int_{-\infty}^{\infty}x(t)e^{-i2\pi ft}\,dt\\
Z(f)&=\int_{-\infty}^{\infty}z(t)e^{-i2\pi ft}\,dt.
\end{align*}
Let $x_T(t)$ be the fraction of the signal $x(t)$ that is equal to $x(t)$ in the time interval $[-\frac{T}{2}, \frac{T}{2}]$ and $0$ outside. The total signal power $P_{\text{tot}}$ is given by 
\begin{equation*}
P_{\text{tot}}=\lim_{T\rightarrow\infty}\frac{1}{T}\int_{-\infty}^{\infty}|x_T(t)|^2\,dt=\lim_{T\rightarrow\infty}\frac{1}{T}\int_{\infty}^{\infty}|X_T(f)|^2\,df=\int_{\infty}^{\infty}P_x(f)\,df
\end{equation*} 
where  
\begin{equation*}
|X_T(f)|^2=\int_{-\infty}^{\infty}\Big[\int_{-\infty}^{\infty}x_T(t-\tau)x_T(t)\,dt\Big]e^{-i2\pi f\tau}\,d\tau
\end{equation*}
 and  $P_x(f)= \lim_{T\rightarrow\infty}\frac{|X_T(f)|^2}{T}$ is the p.s.d. of the signal $x(t)$. Similarly, let $z_T(t)$ be equal to the noise $z(t)$ in the time interval $[-\frac{T}{2}, \frac{T}{2}]$ and $0$ outside. The noise p.s.d. is given by 
\begin{equation*}
	N(f)=\int_{-\infty}^{\infty}R_{z}(\tau)e^{-i2\pi f\tau}\,d\tau
\end{equation*}
with $R_{z}(\tau)=\lim_{T\rightarrow\infty}\frac{1}{T}\int_{-\infty}^{\infty}z_T(t-\tau)z_T(t)\,dt$.

We consider only band-limited signals. Letting the bandwidth be $B>0$, the p.s.d. of band-limited signals with bandwidth $B$ has the following structure:
\begin{equation*}
	P_x(f)=\begin{cases}
	P_x(f) &\text{for } f\in [0,B]\\
		0 &\text{else}.
	\end{cases}
\end{equation*}

$z(t)$ is a band-limited colored Gaussian noise with spectral density $N(f)$ with
\begin{equation*}
	N(f) = \begin{cases}
		\geq 0 &\text{for } f\in [0,B]\\
		0 & \text{else}.
	\end{cases}
\end{equation*}

We consider a communication scenario subject to a power constraint $P$. This means that the total signal power should not exceed $P$, and it is described by 
\begin{equation*}
	\int_{0}^B P_x(f)\,df\leq P.
\end{equation*}

We aim to find codes for the band-limited channel described above. The code should consist of band-limited signals. For this we consider the set $\sX(B,T,P)$ which is  the set of approximately band-limited signals with bandwidth $B$, approximately time-limited to $T$ seconds and with a total power not exceeding $P$, i.e., for every signal $x\in\sX(B,T,P)$ it holds that $\int_{0}^B P_x(f)\,df\leq P$. We define $\sY(B,T)$ to be the set of received signals, which are approximately band-limited with bandwidth $B$ and approximately time-limited to $T$.

\tcr{A code for the band-limited ACGN channel with power constraint consists of a pair of functions $(f,\phi)$, where $f$ is an encoder function $f\colon\sM\rightarrow \sC \subset\sX(B,T,P)$, where $\sC$ is the codebook, and a decoder function $\phi\colon\sY(B,T)\rightarrow \sM$. The transmission rate $R$ is defined by 
\begin{equation*}
	R=\frac{1}{T}\ln |\sM|. 
\end{equation*}
The average error probability $P_e$ is given by 
\begin{equation*}
	P_e=\frac{1}{|\sM|}\sum_{i=1}^{|\sM|}\Pr(\phi(f(i))\neq i).
\end{equation*}
A rate $R$ is called \emph{achievable} for the band-limited ACGN channel, if one can find a code $(f,\phi)$ that operates at a transmission rate of $R$ and for which the average error probability vanishes $P_e\rightarrow 0$ as $T\rightarrow\infty$. The channel \emph{capacity} is defined as the supremum of all achievable rates.}

\begin{thm}[\cite{shannon1949communicationnoise}]\label{thm:Shannon}
	\tcr{The capacity of the band-limited ACGN channel with bandwidth $B$, and continuous noise power spectrum $N$ on the interval $[0,B]$ subject to a power constraint $P>0$ is given by }
	\begin{equation*}
		C(N,P)=\int_0^B \ln\Big(1+\frac{\tcb{P_x^*(f)}}{N(f)}\Big)\,df.
	\end{equation*}
The capacity-achieving power spectrum density is given by
	\begin{equation}
	\tcb{P_x^*(f)}=\begin{cases}
	\Big[\nu - N(f)\Big]_+ &\text{ for } f\in[0,B]\\
	0 &\text{f} \notin [0,B],
	\end{cases}
	\end{equation}
	where $\nu$ is chosen such that $\int_0^B\tcb{P_x^*(f)}\,df=P$ is satisfied.
\end{thm}

\tcr{There are a large number of different derivations for the formula, see \cite{cover1999elements, proakis2001digital, tse2005fundamentals, haykin2001communication}. 

\tco{The capacity-achieving p.s.d.} is given by the water pouring solution. Water pouring is well known for its simple derivation \cite{flanagan2006proving}.}  In general, the problem is approached by dividing the noise spectrum into $n$ subchannels of width $\Delta f_n$ and assuming that each subchannel is \tco{independent} of the others. $N(f)$ is then approximated by $N(f_i)$ for $f\in[f_i-\frac{\Delta f_n}{2},f_i+\frac{\Delta f_n}{2} ]$ and $i\in\{1,\dots, n\}$.  The capacity of each sub channel $f_i$ is given by
 	\begin{equation*}
 	C_n(N,P,f_i)=\Delta f_n \ln \Big(1+\frac{\tcb{P_x^*(f_i)}}{N(f_i)}\Big)
 	\end{equation*}
where  $\tcb{P_x^*(f_i)} =\nu-N(f_i)$ and $\nu$ is derived by using the method of Lagrange multipliers. The total capacity and the total trasmit power are given by 
 	\begin{align}
 	C_n(N,P)&= \sum_{i=1}^n \Delta f_n \ln \Big(1+\frac{\tcb{P_x^*(f_i)}}{N(f_i)}\Big)\label{eq:cn}\\
 	P &= \sum_{i=1}^n \Delta f_n \tcb{P_x^*(f_i)}.\label{eq:pn}
 	\end{align}
 As $n\rightarrow\infty$ then $\Delta f_n\rightarrow 0 $ and Eqs. \refeq{eq:cn} and \refeq{eq:pn} become integrals: 
 	\begin{align*}
 	C(N,P)&=\lim_{n\rightarrow \infty}C_n(N,P)= \int_0^B \ln\Big(1+\frac{\tcb{P_x^*(f)}}{N(f)}\Big)\,df\\
 	P &= \lim_{n\rightarrow \infty}\sum_{i=1}^n \Delta f_n \tcb{P_x^*(f_i)}=\int_0^B \tcb{P_x^*(f)}\,df.
 	\end{align*}
 	
 In general, to show a channel capacity result it is necessary to show achievability and converse. The achievability refers to the possibility of asymptotically achieving error free communication at rates less than the capacity, and the converse shows the impossibility of asymptotically achieving error free communication at rates exceeding the capacity.
 
Achievability results give lower bounds on the capacity. To establish the achievability of band-limited ACGN channel, one must demonstrate the possibility of constructing almost band-limited and almost time-limited codebooks that operate at a rate lower than the channel capacity, i.e., $R<C$. For a given error probability $P_e>0$, the achievability provides us with a monotonically increasing sequence of achievable rates $\{R_n\}_{n\in\N}$ that converges to the capacity as the signal duration $\{T_n\}_{n\in\N}$ increases, i.e., $\{T_n\}_{n\in\N}$ is a monotonically increasing sequence of time duration. For $n\in\N$, the rate $R_n$ describes the codebook size of band-limited signals of $T_n$ time duration for which it is possible to find a decoder strategy, such that the error probability does not exceed $P_e$.

 The converse gives an upper bound on the coding theorem. More specifically, a converse provides us with a monotonically decreasing sequence of rates $\{U_n\}_{n\in\N}$ that converges to the capacity. For every $n\in\N$, $U_n$ is an upper bound on the codebook size of band-limited signals of $T_n$ time duration for which it is possible to find a decoder strategy, such that the error probability does not exceed $P_e$. 
 
Finding algorithms that can calculate both lower and upper bounds would be useful. Moreover, it would be desirable to have an algorithm that takes a band-limited ACGN channel and computes its corresponding capacity-achieving code. This prompts the following question:

{\bf Question 1:} \emph{Given a noise spectral density $N$, a power constraint $P$ and a precision $M$, is it possible find an algorithm that takes $N$, $P$, and $M$ as input and computes a codebook and a decoding strategy with rate $R$ for the band-limited ACGN channel, such that $R\geq C(N,P)-\frac{1}{2^M}$ is achieved? } 

  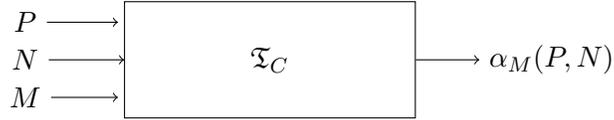
\begin{figure}
   \centering
     \begin{tikzpicture}       [block/.style={draw,minimum width=#1,minimum height=4em},
        block/.default=10em,high/.style={minimum height=3em},auto]

     	\node (b) at (-1,0.5)  {$P$};
       \node (c) at (-1,0) {$N$};
       \node (d) at (-1,-0.5) {$M$};
       \node [block, right=of c] (a) {$\fT_{C}$};
       \node (e) at (6,0) {$\alpha_M(P,N)$};


       \draw[->,draw=black] (b) -- (0.23,0.5);
       \draw[->,draw=black] (c) -- (a);
       \draw[->,draw=black] (d) -- (0.23,-0.5);
       \draw[->,draw=black] (a) -- (e);
            
     \end{tikzpicture}
     
     \caption{Turing machine $\fT_{C}$ for the computation of the capacity approximation of band-limited ACGN channels. It takes the power constraint $P$, noise power spectrum $N$ and the approximation precision $M$ and computes $\alpha_M(P,N)$ with $|C(N,P)-\alpha_M(P,N)|<\frac{1}{2^M}$.}
	\label{fig:TMC}
  \end{figure}

There has been a long-standing interest in the algorithmic computation of the capacity of communication scenarios in information theory. One example of such an algorithm is the Blahut-Arimoto algorithm, which can compute the capacity of any computable discrete memoryless channel given as input (see \cite{arimoto1972algorithm,blahut1972computation}). However, there is no algorithm known to date that can compute the capacity of a band-limited ACGN channel with noise spectral density $N(f)$ similarly to the Blahut-Arimoto algorithm. Ideally, it is desirable to find such an algorithm.

{\bf Question 2:} \emph{Is it possible to find an algorithm that takes a noise spectral density $N$, a power constraint $P$, and a precision $M$ as input and computes the number $\alpha(N,M)$ with 
\begin{equation*}
	|C(N,P)-\alpha(N,M)|<\frac{1}{2^M}?
\end{equation*} } 
The Turing machine that describes the algorithm of Question 2 is illustrated in Fig. \ref{fig:TMC}.

 \begin{figure}
   \centering
     \begin{tikzpicture}       [block/.style={draw,minimum width=#1,minimum height=4em},
        block/.default=10em,high/.style={minimum height=3em},auto]

       \node (c) at (-1,0) {$M$};
       \node [block, right=of c] (a) {$\fT_{C(N,P)}$};
       \node (e) at (6,0) {$\alpha_M$};


       \draw[->,draw=black] (c) -- (a);
       \draw[->,draw=black] (a) -- (e);
            
     \end{tikzpicture}
     
     \caption{Turing machine $\fT_{C(N,P)}$ for the computation of the capacity approximation of band-limited ACGN channels for fixed $N$ and $P$. For $P\in\R_c$ and $N$ computable, $\fT_{(N,P)}$ takes the approximation precision $M$ and computes $\alpha_M$ with $|C(N,P)-\alpha_M|<\frac{1}{2^M}$.}
	\label{fig:TMNP}
  \end{figure}
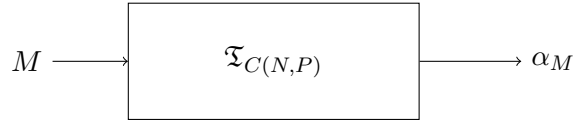

We could simplify the requirements of the desired algorithm by fixing $N$ and $P$. This prompts the following question:

{\bf Question 3:} \emph{For a fixed $N$ and a fixed $P$, is it possible to find an algorithm that takes a precision $M$ as input and computes the number $\alpha(N,M)$ with 
\begin{equation*}
	|C(N,P)-\alpha(N,M)|<\frac{1}{2^M}?
\end{equation*} }
A Turing machine describing the algorithm of Question 3 is illustrated in Fig. \ref{fig:TMNP}.

If a constructive proof is found for Theorem \ref{thm:Shannon}, including an effective construction for capacity-achieving codes and an algorithmic description of the converse, then it would provide a positive answer to all three questions. However, our analysis demonstrates that for fixed $N$ and $P$ it is not possible to provide such a constructive proof.
\section{Computability Framework}
\label{sec:computability_numbers}

In this section, we introduce the fundamental concepts of computability theory, which are needed therefore. Computability and computable real numbers were initially proposed by Turing in \cite{turing1936computable} and \cite{turing1938computable}. In this context, computable numbers denote real numbers that can be computed using Turing machines.

A sequence of rational numbers $\{r_n\}_{n\in\N}$ is called a \emph{computable sequence} if there exist recursive functions $a,b,s:\N\rightarrow\N$ with $b(n)\neq 0$ for all $n\in\N$ and
\begin{equation}
	\label{eq:computability_comp1}
	r_n= (-1)^{s(n)}\frac{a(n)}{b(n)}, \qquad n\in\N.
\end{equation}

A real number $x$ is said to be computable if there exists a computable sequence of rational numbers $\{r_n\}_{n\in\N}$, such that
\begin{equation}
	\label{eq:computability_comp2}
	|x-r_n|<2^{-n}
\end{equation}
for all $n\in\N$. This means that the computable real number $x$ is completely characterized by the recursive functions $a,b,s:\N\rightarrow\N$. It has the representation $(a,b,s)$ which we also write as $x\sim (a,b,s)$. It is clear that this representation must not be unique and that there might be other recursive functions $a',b',s':\N\rightarrow\N$ which characterize $x$, i.e., $x\sim (a',b',s')$.

We denote the set of computable real numbers by $\R_c$ and the set of positive computable real numbers by $\R^{\geq 0}_c$. 

\begin{defn}
A sequence $\{r_n\}_{n\in\N}$ of rational numbers converges \emph{effectively} to a real number $x$ if there exists a recursive function $e\colon\N\rightarrow\N$ such that for all $N\in\N$ it holds that
	\begin{equation*}
		k\geq e(N) \quad \text{implies}\quad |r_k-x|\leq 2^k.
	\end{equation*}
\end{defn}

\begin{defn}
	A sequence $\{x_n\}_{n\in\N}$ is called a \emph{Cauchy sequence} if for every $\epsilon>0$, there is a $n_0\in\N$ such that for every $m,n>n_0$ it holds that
	\begin{equation*}
		|x_n-x_m|<\epsilon.
	\end{equation*}
\end{defn}

\begin{defn}
	A Cauchy sequence $\{x_n\}_{n\in\N}$ is said to converge \emph{effective} if  there is a recursive function $e\colon\N\times\N\rightarrow\N$ such that for all $n,N\in\N$ it holds that
	\begin{equation*}
	k\geq e(n,N) \quad \text{ implies } |x_k-x_n|\leq 2^{-N}
	\end{equation*}
\end{defn}

\begin{defn}
	\label{def:borel}
	A function $f_c:\R_c\rightarrow\R_c$ is called \emph{Borel-Turing computable} if there is an algorithm (or Turing machine) that transforms each given representation $(a,b,s)$ of a computable real number $x$ into a corresponding representation for the computable real number $f_c(x)$.
\end{defn}

\begin{defn}
	\label{def:computableseq}
	A sequece of real numbers $\{x_n\}_{n\in\N}$ is \emph{computable} (as a sequence) if there is a computable double sequence of rationals $\{r_{m,n}\}_{(m,n)\in\N^2}$ such that 
	\begin{equation}
		|r_{m,n}-x_n|\leq 2^{-m}
	\end{equation}
	for all $m\in\N$ and $n\in\N$.
\end{defn}

\begin{rem}
Let $\{x_n\}_{n\in\N}$ and $\{y_n\}_{n\in\N}$ be computable sequences of real number. \tco{Then the} following sequence are also computable:
	\begin{equation*}
		x_n\pm y_n,\text{ }x_n y_n,\text{ } x_n/ y_n\text{ } (y_n\neq 0 \text{ for all } n),\text{ } \exp x_n,\text{ } \log x_n \text{ }(x_n>0 \text{ for all } n).
	\end{equation*}
\end{rem}

\begin{defn}[\cite{pour2017computability}]
	\label{def:compcont}
	Let $\mathbb{I}_c\subset\R_c$ be a computable interval. A function $f_c:\mathbb{I}_c\rightarrow\R_c$ is called  \emph{computable continuous} if
	\begin{enumerate}
		\item $f_c$ is \emph{sequentially computable}, i.e., $f_c$ maps every computable sequence $\{x_n\}_{n\in\N}$ of points $x_n\in\mathbb{I}_c$ into a computable sequence $\{f_c(x_n)\}_{n\in\N}$ of real numbers, and
		\item $f_c$ is \emph{effectively uniformly continuous}, i.e., there is a recursive function $d:\N\rightarrow\N$ such that for all $x,y\in\mathbb{I}_c$ and all $N\in\N$ with
		\begin{equation*}
			\|x-y\|\leq\frac{1}{d(N)}
		\end{equation*}	
		it holds that
		\begin{equation*}
			|f_c(x)-f_c(y)|\leq\frac{1}{2^N}.
		\end{equation*}
	\end{enumerate}
\end{defn}
\begin{rem}
The notion of computable continuous functions is stronger than that of Borel-Turing computable functions. Functions that are computable continuous are also Borel-Turing computable.
\end{rem}

\begin{lem}[\cite{pour2017computability}]
Let $[a,b]\subset \R_c$. Let $f\colon [a,b]\rightarrow \R$ be a computable function. Then the definite integral 
	\begin{equation}
		\tcr{v=\int_a^b f(x)\,dx}
	\end{equation}
	is a computable real number.
\end{lem}
We further need the concepts of a recursive set and a recursively enumerable set as defined in \cite{soare1978recursively}. These are used with the purpose of constructing sequences of computable channels used to study the computability of the feedback capacity function.

\begin{defn}
	\label{def:recursive}
	A set $\sA\subset\N$ is called \emph{recursive} if there exists a computable function $f$ such that $f(x)=1$ if $x\in\sA$ and $f(x)=0$ if $x\notin\sA$. 
\end{defn}

\begin{defn}
	\label{def:recursiveenumerable}
	A set $\sA\subset\N$ is \emph{recursively enumerable} if there exists a recursive function whose domain is exactly $\sA$.
\end{defn}

We have the following properties \cite{soare1978recursively}:
\begin{itemize}
	\item $\sA$ is recursive is equivalent to: $\sA$ is recursively enumerable and $\sA^c$ is recursively enumerable.
	\item There exist recursively enumerable sets $\sA\subset\N$ that are not recursive, i.e., $\sA^c$ is not recursively enumerable. This means there are no computable, i.e., recursive, functions $f:\N\rightarrow\sA^c$ with $[f(\N)]=\{m\in\N\colon \exists n \in\N\;\; \text{with}\;\; f(n)=m\}=\sA^c$.
\end{itemize}
%


\section{Computability of the ACGN Channel Capacity}
\label{sec:computability_Gauss}

In this section, we aim to address Question 1, Question 2, and Question 3 from Section \ref{sec:Gauss_Channel}. Specifically, we construct an example of a noise power spectrum $N$ for a band-limited ACGN channel that yields a negative answer to all three questions. To achieve this, we will introduce a band-limited ACGN channel that has computable parameters, including computable bandwidth, computable noise power spectrum, computable capacity-achieving power density spectrum, and a computable power constraint. Both the noise power spectrum and the capacity-achieving power spectrum will be computable continuous functions of the frequency domain $f\in\mathbb{R}$.

We consider the following band-limited channel with bandwidth $B\in\R_c$ and $B>0$
	\begin{equation*}
	y(t) = x(t)+z(t).
	\end{equation*} 
$N(f)$ is also band-limited with a $B$ bandwidth. For $f\in[0,\frac{B}{2}]$, $N(f)$ is strictly monotonically decreasing and for $f\in[\frac{B}{2},B]$ is strictly monotonically increasing. $N$ is an even function with respect to $\frac{B}{2}$.

The communication is subject to a power constraint $P\in\R_c$, $P>0$. The p.s.d. $P_x(f)$ with $f \in [0,B]$ is a non-negative continuous function with
	\begin{equation*}
		\int_{0}^BP_x(f)\,df=P
	\end{equation*}

We denote the capacity achieving p.s.d. by $P^*_x(f)$ , which is uniquely determined by the water pouring technique.

We choose $f_1\in\Big(0,\frac{B}{2}\Big]$, $f_1\in\R_c$. We want to look for a capacity-achieving p.s.d. $\tcb{P_{f_1}^*(f)}$, that is different from zero only in the interval $\big[\frac{B}{2}-f_1,\frac{B}{2}+f_1\big]$. This optimal p.s.d. is uniquely coupled with the power $\tcb{P_{f_1}}$ and is given by
	\begin{equation*}
		\tcb{P_{f_1}}=\int_{\frac{B}{2}-f_1}^{\frac{B}{2}+f_1}\Big(N\Big(\frac{B}{2}+f_1\Big)-N(f)\Big)\,df.
	\end{equation*}
 Note that $N(\frac{B}{2}+f_1) = N\Big(\frac{B}{2}-f_1\Big)$. $\tcb{P_{f_1}}$ is a computable number.

This means that when we are given a power $P$, such that for a certain $\hat{f_1}$ we have that $P=\tcb{P_{\hat{f_1}}}$, we have that the optimal power allocation is
	\begin{equation*}
		\tcb{P_{\hat{f_1}}^*(f)}=
		\begin{cases}
			N\Big(\frac{B}{2}+\hat{f_1}\Big)-N(f), &\text{for }f\in\Big[\frac{B}{2}+\hat{f_1},\frac{B}{2}-\hat{f_1}\Big]\\
			0, &\text{otherwise.}
		\end{cases}						
	\end{equation*}	

The corresponding capacity can hence be expressed as a function of $f_1$ and is given by 
	\begin{equation*}
		C_1(N,f_1)=\int_0^B \ln (\tcb{P_{f_1}^*(f)}+N(f))\,df - \int_0^B \ln(N(f))\,df.
	\end{equation*}
Now, if $C(N,P)$ is the capacity of the band-limited ACGN channel, then the following representation applies for $0\leq f_1\leq \frac{B}{2}$ and the corresponding power $\tcb{P_{f_1}}$:
	\begin{equation*}
		C(N,\tcb{P_{f_1}})=C_1(N,f_1),
	\end{equation*}
i.e., the capacity is a function of $f_1$.


\begin{thm}\label{thm:capacitycomp}
Let $B>0$, $B\in\R_c$ be arbitrary. There are computable continuous functions $N\colon [0,B]\rightarrow \R^{\geq 0}_c$, such that for all $f_1\in [0,\frac{B}{2}]$, $f_1\in\R_c$ we have that
	\begin{equation*}
		\tco{C_1(N,f_1)\notin\R_c.}
	\end{equation*}
Moreover, for every $f_1\in[0,\frac{B}{2}]$, $f_1\in\R_c$, there is no computable sequence of computable numbers $\{u_n\}_{n\in\N}$ with $u_n\geq u_{n+1}$, $n\in\N$ and 
	\begin{equation*}
		\lim_{n\rightarrow\infty}u_n = C_1(N,f_1).
	\end{equation*}
\end{thm}

\begin{proof}
Here we prove the result of Theorem \ref{thm:capacitycomp}. For this, we construct a  non-negative computable continuous p.s.d. $N$. The construction of $N$ is based on a recursively enumerable non-recursive set $\sA$. There are countably infinitely many recursively enumerable non-recursive sets \cite{soare1978recursively}. We denote $\{\sA_i\}_{i\in\N}$ as the family of recursively enumerable non-recursive sets.  For every such set $\sA_i$, one can use the same approach to construct a different non-negative computable continuous p.s.d. $N_i$. The capacity of every $N_i$ yields a non-computable number $\xi_i$. 

Next, we start with the construction of the noise p.s.d. $N$. Let $B>0$ be a fixed computable number. Let $n_0\in\N$, such that $\frac{1}{n_0}<\frac{B}{2}$.

We consider the following function for $n\geq n_0$

 	\begin{equation}
		G_n(f)=
			\begin{cases}
				-\frac{1}{|f-\frac{B}{2}|}, &\text{for } f\in\Big[0,\frac{B}{2}-\frac{1}{n}\Big]\cup\Big[\frac{B}{2}+\frac{1}{n}, \frac{B}{2}\Big]\\
				-n, &\text{for } f\in\Big[\frac{B}{2}-\frac{1}{n},\frac{B}{2}+\frac{1}{n}\Big]\\
			\end{cases}
	\end{equation}

$G_n$ is a computable continuous function. Let 
	
	\begin{align}
		C_n &= \int_0^B G_n(f)\, df = 2 \int_{\frac{B}{2}}^B G_n(f)\, df\nonumber \\
			&= -2 \int_0^{\frac{1}{n}}n\,df - \int_{\frac{1}{n}}^{\frac{B}{2}} \frac{1}{f}\, df\nonumber\\
			&= -2 -2\Big(\log\frac{B}{2}-2\ln\frac{1}{n}\Big)\nonumber\\
			&= -2\Big(1+\ln\frac{nB}{2}\Big).\label{eq:Cn}
	\end{align}
	
Note that $\frac{nB}{2}>1$, and hence $\ln\frac{nB}{2}>0$.

We set $C_n^{(1)}=|C_n|$.

Let $\sA\in\N$ be a recursively enumerable non-recursive set. Let $\varphi_\sA\colon \N\rightarrow\sA$ be a recursive function that lists all elements of the set $\sA$.

We consider the following sequence of functions:

	\begin{equation}\label{eq:N}
		N_M(f)= \Big(f-\frac{B}{2}\Big)^2 \exp\Big(\sum_{n=1}^M \frac{1}{2^{\varphi_\sA(n)}}\frac{1}{C_n^{(1)}}G_n(f)\Big)
	\end{equation}

$N_M$ is a computable continuous function on $[0,B]$, since $G_n$ are computable continuous functions for $1\leq n \leq M$, the exponential function $\exp(\cdot)$ maps computable continuous functions to computable continuous functions, and the multiplication with $\Big(f-\frac{B}{2}\Big)^2$ generates, in any case, computable functions. Hence, $\{N_M\}_{M\in\N}$ is a computable sequence of computable continuous functions.
$N_M$ is itself a strictly monotonically increasing function in $[\frac{B}{2},B]$ and an even function with respect to $\frac{B}{2}$.

Let $K\in\N$ be arbitrary. We have 
	\begin{align*}
		N_{M+K}(f)-N_M(f) = \Big(f-\frac{B}{2}\Big)^2 \exp\Big(\sum_{n=1}^M \frac{1}{2^{\varphi_\sA(n)}}\frac{1}{C_n^{(1)}}G_n(f)\Big)\Big[ \exp\Big(\sum_{n=M+1}^{M+K} \frac{1}{2^{\varphi_\sA(n)}}\frac{1}{C_n^{(1)}}G_n(f)\Big)-1\Big],
	\end{align*}

	\begin{align*}
		|N_{M+K}(f)-N_M(f)| = \Big(f-\frac{B}{2}\Big)^2 \exp\Big(\sum_{n=1}^M \frac{1}{2^{\varphi_\sA(n)}}\frac{1}{C_n^{(1)}}G_n(f)\Big)\Big|1- \exp\Big(\sum_{n=M+1}^{M+K} \frac{1}{2^{\varphi_\sA(n)}}\frac{1}{C_n^{(1)}}G_n(f)\Big)\Big|.
	\end{align*}

For $x\in[0,1]$ we have
	\begin{equation*}
		1-e^{-x}\leq 2x.
	\end{equation*}

Let $L\in\N$ with $L>n_0$ be arbitrary. On the interval $[0,\frac{B}{2}-\frac{1}{L}]$ and $[\frac{B}{2}+\frac{1}{L}, B]$ we have that for $M>L$
	\begin{equation*}
		\frac{B}{2}-\frac{1}{M}>\frac{B}{2}-\frac{1}{L} \quad\text{and}\quad \frac{B}{2}+\frac{1}{M}<\frac{B}{2}+\frac{1}{L}.
	\end{equation*}
Hence 
	\begin{equation*}
		G_M(f)=-\frac{1}{|f-\frac{B}{2}|} \quad\text{for}\quad f\in \Big[0,\frac{B}{2}-\frac{1}{L}\Big]\cup \Big[\frac{B}{2}+\frac{1}{L}, B\Big],
	\end{equation*}
so that for $M>L$ the following holds:
	\begin{equation*}
		\sum_{n=M+1}^{M+K} \frac{1}{2^{\varphi_\sA(n)}}\frac{1}{C_n^{(1)}}G_n(f)= -\frac{1}{|f-\frac{B}{2}|}\sum_{n=M+1}^{M+K} \frac{1}{2^{\varphi_\sA(n)}}\frac{1}{C_n^{(1)}}
	\end{equation*}
and

	\begin{align*}
		\sum_{n=M+1}^{M+K} \frac{1}{2^{\varphi_\sA(n)}}\frac{1}{C_n^{(1)}}&<\frac{1}{C_{M+1}^{(1)}}\sum_{n=M+1}^{M+K} \frac{1}{2^{\varphi_\sA(n)}}\\
		&<\frac{1}{C_{M+1}^{(1)}}\sum_{n=1}^{\infty} \frac{1}{2^{\varphi_\sA(n)}}\\
		&< \frac{1}{C_{M+1}^{(1)}}\sum_{n=1}^{\infty} \frac{1}{2^{n}}\\
		&= \frac{1}{C_{M+1}^{(1)}}.
	\end{align*}
Here we have used that $\{C_{n}^{(1)}\}_{n\in\N}$ is a monotonically increasing sequence. 
It also holds that 
	\begin{align*}
		0&\geq \sum_{n=M+1}^{M+K} \frac{1}{2^{\varphi_\sA(n)}}\frac{1}{C_n^{(1)}}G_n(f)\\
		&\geq -\frac{1}{|f-\frac{B}{2}|} \frac{1}{C_{M+1}^{(1)}}
	\end{align*}
We also have that
	\begin{equation*}
		\frac{1}{|f-\frac{B}{2}|}\leq \frac{1}{|\frac{B}{2}+\frac{1}{L}-\frac{B}{2}|}=L,
	\end{equation*}
hence
	\begin{equation}\label{eq:A2}
		0\geq  \sum_{n=M+1}^{M+K} \frac{1}{2^{\varphi_\sA(n)}}\frac{1}{C_n^{(1)}}G_n(f)\geq - \frac{L}{C_{M+1}^{(1)}}.
	\end{equation}

For $M\in\N$ and $K\in\N$, such that $M>U3^{L^2}$ with $U\in\N$, $U\geq 1$, $\frac{B}{2}U>1$ we have that 

	\begin{align}
		|N_{M+K}(f)-N_M(f)| &\leq \Big(f-\frac{B}{2}\Big)^2 \exp\Big(\sum_{n=1}^M \frac{1}{2^{\varphi_\sA(n)}}\frac{L}{C_n^{(1)}}G_n(f)\Big)\frac{L}{C_{M+1}^{(1)}}\nonumber\\
		&< \Big(\frac{B}{2}\Big)^2\times 1 \times \frac{L}{2(1+\ln\frac{BU3^{L^2}}{2})}\label{eq:ineq}\\
		&=\Big(\frac{B}{2}\Big)^2\times 1 \times \frac{L}{(2+2\log\frac{BU}{2}+2\ln 3^{L^2})}\nonumber\\
		&<\Big(\frac{B}{2}\Big)^2\frac{L}{(2\ln 3^{L^2})}\nonumber\\
		&<\Big(\frac{B}{2}\Big)^2\frac{L}{2L^2}\label{eq:A3}.
	\end{align}
	
	\eqref{eq:ineq} follows from \eqref{eq:Cn}.
	
	Let $U_1$ be the smallest natural number such that $U_1>\frac{B}{2}$. Then for all $L>n_0$, $L\in\N$, for all $M\geq U3^{L^2}$, for all $K\in\N$, and for all $f\in[0,\frac{B}{2}-\frac{1}{L}]\cup[\frac{B}{2}+\frac{1}{L},B]$, we have that 
	\begin{equation*}
		|N_{M+K}(f)-N_M(f)| <\frac{U_1^2}{2L}.
	\end{equation*}
	
	For $f\in[\frac{B}{2}-\frac{1}{L},\frac{B}{2}+\frac{1}{L}]$ we have the following:
	
	\begin{align*}
		|N_{M+K}(f)-N_M(f)|&\leq  \Big(f-\frac{B}{2}\Big)^2 \Big[\exp\Big(\sum_{n=1}^{M+K} \frac{1}{2^{\varphi_\sA(n)}}\frac{1}{C_n^{(1)}}G_n(f)\Big)+\exp\Big(\sum_{n=1}^{M} \frac{1}{2^{\varphi_\sA(n)}}\frac{1}{C_n^{(1)}}G_n(f)\Big)\Big]\\
		&\leq 2\Big(f-\frac{B}{2}\Big)^2\\
		&\leq 2\Big(\frac{B}{2}+\frac{1}{L}-\frac{B}{2}\Big)^2\\
		&\leq \frac{1}{2L^2}.
	\end{align*}
Thus $\{N_M\}_{M\in\N}$ is an effective Cauchy sequence of computable continuous functions, and it converges effectively to the function 
	\begin{equation}\label{eq:n}
		N(f)=\Big(f-\frac{B}{2}\Big)^2\exp\Big(\sum_{n=1}^{\infty} \frac{1}{2^{\varphi_\sA(n)}}\frac{1}{C_n^{(1)}}G_n(f)\Big).
	\end{equation}
	
Eq. \eqref{eq:n} describes an algorithm that takes the recursive function $\varphi_{\sA}$ as input and computes $N$. $N$ is a computable continuous function with $N(f)\geq 0$ for $f\in[0,B]$. $N$ is a strictly monotonically increasing function in the interval $[\frac{B}{2},B]$ and it is an even function with respect to $\frac{B}{2}$.

We take an $f_1\in(0,B]$, $f_1\in\R_c$, and compute the number $C_1(N,f_1)$.
For 
	\begin{equation*}
		\tcb{P_{f_1}}=\int_{\frac{B}{2}-f_1}^{\frac{B}{2}+f_1}\Big(N\Big(\frac{B}{2}+f_1\Big)-N(f)\Big)\,df
	\end{equation*}
we have that
	\begin{align}
		C_1(N,f_1)&= \int_{\frac{B}{2}-f_1}^{\frac{B}{2}+f_1}\ln(\tcb{P_{f_1}^*(f)}+N(f))\,df +\int_{0}^{\frac{B}{2}-f_1}\ln(N(f))\,df\nonumber\\
		&\quad +\int_{\frac{B}{2}+f_1}^{B}\ln(N(f))\,df- \int_{0}^{B}\ln(N(f))\,df\nonumber\\
		&= \int_{\frac{B}{2}-f_1}^{\frac{B}{2}+f_1}\ln\Big(N\Big(\frac{B}{2}+f_1\Big)\Big)\,df_1-\int_{\frac{B}{2}-f_1}^{\frac{B}{2}+f_1}\ln(N(f))\,df.\label{eq:A4}
	\end{align}
Since $f_1\in\R_c$, we have that $\frac{B}{2}+f_1\in\R_c$ and hence $N(\frac{B}{2}+f_1)\in\R_c$ and $N(\frac{B}{2}+f_1)>0$. Consequently, we have that $\ln(N(\frac{B}{2}+f_1))\in\R_c$ and therefore also $2f_1\ln N(\frac{B}{2}+f_1)\in\R_c$. 

Now we have to rewrite the number
	\begin{align}
		Z(f_1)&= \int_{\frac{B}{2}-f_1}^{\frac{B}{2}+f_1}\ln N(f)\,df\nonumber\\
		&=  \int_{0}^{B}\ln N(f)\,df- \int_{0}^{\frac{B}{2}-f_1}\ln N(f)\,df- \int_{\frac{B}{2}+f_1}^{B}\ln N(f)\,df\label{eq:A3}
	\end{align}
	 We then have
	 \begin{align*}
		\int_{0}^{B}\ln N(f)\,df&= \int_{0}^{B}\ln \Big(f-\frac{B}{2}\Big)^2\,df + \int_{0}^{B}\ln\Big(\exp\Big(\sum_{n=1}^{\infty} \frac{1}{2^{\varphi_\sA(n)}}\frac{1}{C_n^{(1)}}G_n(f)\Big)\Big)\,df
	\end{align*}
	
	We have that $\ln \Big(\cdot-\frac{B}{2}\Big)^2$ is a computable function in $\mathcal{L}^1[0,B]$, see \cite{pour2017computability}.
	This way, we have that 
	\begin{equation}
		\int_0^B\ln \Big(\cdot-\frac{B}{2}\Big)^2\,df\in\R_c.
	\end{equation}
	Furthermore, we also have that
	\begin{equation*}
		\exp\Big(\sum_{n=1}^{\infty} \frac{1}{2^{\varphi_\sA(n)}}\frac{1}{C_n^{(1)}}G_n(f)\Big)=2^{\log_2 \exp\Big(\sum_{n=1}^{\infty} \frac{1}{2^{\varphi_\sA(n)}}\frac{1}{C_n^{(1)}}G_n(f)\Big) }
	\end{equation*}
	and
	\begin{equation*}
		\int_0^B\log_2\Big(\exp\Big(\sum_{n=1}^{\infty} \frac{1}{2^{\varphi_\sA(n)}}\frac{1}{C_n^{(1)}}G_n(f)\Big)\Big)\,df = \log_2(e)\int_0^B \sum_{n=1}^{\infty} \frac{1}{2^{\varphi_\sA(n)}}\frac{1}{C_n^{(1)}}G_n(f)\,df.
	\end{equation*}
	We consider the following function for $f\in[0,B]$ and $M\in\N$:
	\begin{equation*}
		\psi_M(f)= \sum_{n=1}^{M} \frac{1}{2^{\varphi_\sA(n)}}\frac{1}{C_n^{(1)}}G_n(f).
	\end{equation*}
	Note that $\psi$ is a continuous function.
	For $K\in\N,$ it holds that 
	\begin{align}
		\int_0^B |\psi_{M+K}(f)-\psi_{M}(f)|\,df &= \int_0^M \sum_{n=M}^{M+K} \Big|\frac{1}{2^{\varphi_\sA(n)}}\frac{1}{C_n^{(1)}}G_n(f)\Big|\,df\nonumber\\
		&\leq \sum_{n=M}^{M+K} \frac{1}{2^{\varphi_\sA(n)}}\frac{1}{C_n^{(1)}}\int_0^M |G_n(f)|\,df\label{eq:un3eck}\\
		&= \sum_{n=M}^{M+K} \frac{1}{2^{\varphi_\sA(n)}}\label{eq:defC}\\
		&< \sum_{n=M}^{\infty} \frac{1}{2^{\varphi_\sA(n)}}\nonumber.
	\end{align}
	Eq. \eqref{eq:un3eck} holds due to the triangle inequality for the $\ell_1-$norm. Eq. \eqref{eq:defC} holds from the definition of $C_n^{(1)}$. 
	Consequently, the sequence $\{\psi_{M}(f)\}_{M\in\N}$ converges in the $\ell_1-$norm to the function
	\begin{equation*}
		\psi(f)= \sum_{n=M}^{\infty} \frac{1}{2^{\varphi_\sA(n)}}\frac{1}{C_n^{(1)}}G_n(f)
	\end{equation*}
	for $f\in [0,B]$.
	
	\tco{Thus, we have that} 
	\begin{align*}
		\int_0^B\psi(f)\,df &=\lim_{M\rightarrow \infty}\int_0^B \sum_{n=1}^{M} \frac{1}{2^{\varphi_\sA(n)}}\frac{1}{C_n^{(1)}}G_n(f)\,df\\
		&= \lim_{M\rightarrow \infty} \sum_{n=0}^{M} \frac{1}{2^{\varphi_\sA(n)}}\frac{1}{C_n^{(1)}}\int_0^BG_n(f)\,df\\
		&= \lim_{M\rightarrow \infty} -\sum_{n=0}^{M} \frac{1}{2^{\varphi_\sA(n)}}\\
		&=  -\sum_{n=0}^{\infty} \frac{1}{2^{\varphi_\sA(n)}}.
	\end{align*}
Since $\sA\subset \N$ is a recursively enumerable non-recursive set and from \cite[Chapter~1]{pour2017computability}, we have
	\begin{equation*}
		-\sum_{n=0}^{\infty} \frac{1}{2^{\varphi_\sA(n)}}= \xi \notin\R_c.
  	\end{equation*}		

Next, we must analyze the integral $\int_0^{\frac{B}{2}-f_1}\log N(f)\,df$.
We have already shown in the inequality \eqref{eq:A2} that for $L\in\N$ with $\frac{1}{L}<f_1$, it always holds the following relation for $M>U3^{L^2}$:
	\begin{equation*}
		|N_{M+K}(f)-N_M(f)|<\frac{L}{C_{M+1}^{(1)}}.
	\end{equation*}
Following similar calculations as in Eq. \eqref{eq:A3}, for $f\in[0,\frac{B}{2}-f_1]$ and $M>U3^{L^2}$ we have that
	\begin{equation*}
		|\psi_{M+K}(f)-\psi_M(f)|<\frac{1}{L}.
	\end{equation*}
	
Since $\{\psi_{M}(f)\}_{M\in\N}$ is a computable continuous sequence of continuous functions on $[0,\frac{B}{2}-f_1]$. This sequence converges effectively on $[0,\frac{B}{2}-f_1]$ to the function $\psi$. $\psi$ is itself a computable continuous function on $[0,\frac{B}{2}-f_1]$. With this  and from \cite{pour2017computability}, it follows that
	\begin{equation}\label{eq:A31}
		\int_0^{\frac{B}{2}-f_1}\psi(f)\,df \in\R_c.
	\end{equation}
	
Following the same line of arguments, we get that $\psi$ is also a computable continuous function on the interval $[\frac{B}{2}+f_1,B]$, and hence
	\begin{equation*}\label{eq:A32}
		\int_{\frac{B}{2}+f_1}^B\psi(f)\,df \in \R_c.
	\end{equation*}
	
From Eqs. \eqref{eq:A3}, \eqref{eq:A31} and \eqref{eq:A32} it must hold that
	\begin{equation*}
		Z(f_1)\notin\R_c,
	\end{equation*}
and hence
	\begin{equation*}
		C_1(N,f_1)\notin\R_c.
	\end{equation*}
	
	Since $f_1\in\R_c$ can take any value in the interval $(0,\frac{B}{2}]$, we have shown the \tco{first} statement of Theorem \ref{thm:capacitycomp}.
	
	Next, we show the second statement. We prove this by contradiction and assume that the second statement is wrong.
	Assume that there is a $f_1\in(0,\frac{B}{2}]$, $f_1\in\R_c$ so that we can find a computable sequence of computable numbers $\{u_n\}_{n\in\N}$, such that the following holds:
	\begin{equation*}
		u_n\geq u_{n+1} \text{ for } n\in\N \quad \text{ and }\lim_{n\rightarrow \infty}u_n= C_1(N,\hat{f_1}).
	\end{equation*}
	 From the proof of the first statement, we have that
	 \begin{equation*}
	 	C_1(N,\hat{f_1})= a(\hat{f_1}) + \xi
	 \end{equation*}
with $a(\hat{f_1})\in\R_c$ and $\xi = -\sum_{n=1}^\infty \frac{1}{2^{\varphi_{\sA}(n)}}\notin\R_c$.

We set $U(M)=-\sum_{n=1}^M \frac{1}{2^{\varphi_{\sA}(n)}}$ for $M\in\N$. Then we have a computable sequence of computable numbers with 
	\begin{equation*}
		U(M)\geq U(M+1)  \quad \text{ and }\lim_{M\rightarrow \infty}U(M)= \xi.
	\end{equation*}
	
	If $C_1(N,f_1)$ were the limit value of a monotonically decreasing sequence of computable numbers, then this would also hold for $C_1(N,f_1)-a(\hat{f_1})$. Since $a(\hat{f_1})$ is a computable number, there is a monotonically increasing computable sequence $\{m_n\}_{n\in\N}$ of computable numbers with $\lim_{n\rightarrow \infty} m_n=a(\hat{f_1})$. Furthermore, since $a(\hat{f_1})\geq m_n$ it also holds that $-a(\hat{f_1})\leq- m_n$. We then have 
	\begin{equation*}
		u_n-m_n\geq C_1(N,\hat{f_1})- a(\hat{f_1}) \quad n\in\N
	\end{equation*}
	and 
	\begin{equation*}
		u_{n+1}-m_{n+1}\leq u_n-m_{n+1}\leq u_n-m_n,
	\end{equation*}
i.e.,$\{u_n-m_n\}_{n\in\N}$ is a computable sequence of computable numbers and the sequence is monotonically decreasing. It then holds that
	\begin{equation}
		\lim_{n\rightarrow\infty}(u_n-m_n) = \lim_{n\rightarrow\infty}u_n-\lim_{n\rightarrow\infty}m_n=  C_1(N,\hat{f_1})- a(\hat{f_1})=\xi.
	\end{equation}
	
	This way is $\xi$ the limit value of computable sequences. One of the computable sequences is a monotonically decreasing sequence and the other one is a monotonically increasing sequences. This automatically implies that $\xi\in\R_c$ which is a contradiction. This contradiction shows that our assumption is wrong and hence there is no monotonically decreasing computable sequence of computable numbers that converges to $C_1(N,\hat{f_1})$.
\end{proof}

\begin{rem}
Theorem \ref{thm:capacitycomp} states that there are band-limited ACGN channels whose capacities are non-computable numbers. This result is the second known instance in information theory where capacity has been proven to be non-computable, following the compound channel case in \cite{boche2020communication}. In that study, the authors considered a computable compound channel $\{W_n\}_{n\in\N}$ with finite input and output alphabets. $\{W_n\}_{n\in\N}$ is a computable sequence, and $C(\{W_n\}_{n\in\N})\notin R_c$. The capacity $C(\{W_n\}_{n\in\N})$ is the limit value of a monotonically decreasing computable sequence $\{u_n\}_{n\in\N}$ that serves as a computable upper bound for $C(\{W_n\}_{n\in\N})$, but there exists no computable sequence of lower bounds that converges to the capacity. In contrast, Theorem \ref{thm:capacitycomp} shows that for $C(N,P)$, the capacity of the band-limited ACGN channel with colored noise, we have the opposite situation. 
\end{rem}

\begin{rem}
\tco{In previous works such as \cite{shannon1949communicationnoise, gallager1968information, cover1999elements}, the capacity of the ACGN channel was typically related to the capacity of the discrete Gaussian channel. This was achieved through a discrete approximation of the time-continuous Gaussian channel. When analyzing these solutions, it is observed that as the approximation of the discrete channels becomes finer, the sequence of capacities of the discrete channels approaches the capacity of the time-continuous ACGN channel. However, a stopping criterion for the approximation process has not yet been identified. In our case, such a stopping criterion refers to an algorithm that takes an approximation error of $\frac{1}{2^M}$ as input for the computation of the capacity of a fixed ACGN channel, and then the algorithm stops the approximation process when the result of the computation is within a margin of error of $\frac{1}{2^M}$ from the capacity of the time-continuous ACGN channel. Our result shows that there are band-limited ACGN channels with color noise for which such a stopping criterion cannot exist.}
\end{rem}

\begin{rem}
We not only demonstrate the existence of a non-negative computable continuous noise spectral density, but we also develop an algorithm that can effectively construct a noise power spectrum $N$ for which the conclusion of Theorem \ref{thm:capacitycomp} holds. The algorithm takes a recursive function $\varphi_{\sA}$ as input and computes $N$. The recursive function $\varphi_{\sA}$ generates a recursively enumerable non-recursive set $\sA$. There are countably infinitely many recursive enumerable non-recursive sets $\{\sA_1,\sA_2,\dots\}$. By applying the same algorithm to the generative function $\varphi_{\sA_i}$ of any other recursively enumerable non-recursive set $\sA_i$, we obtain a different computable noise power spectrum $N_i$ which has the same structure as $N$ and that satisfies Theorem \ref{thm:capacitycomp}.
\end{rem}

\begin{thm}\label{thm:cardinalityP}
Let $B>0$, $B\in\R_c$ be arbitrary. There are computable continuous noise spectral densities $N\colon[0,B]\rightarrow \R_{\geq0}^{c}$ such that there are infinitely many computable $\hat{P}\in[0,P_*]$ where
	\begin{equation*}
		P_*=\int_0^B(N(0)-N(f))\, df, 
	\end{equation*}
such that $\hat{P}\in\R_c$ \tco{but}
	\begin{equation*}
		\tco{C(N,\hat{P})\notin \R_c}.
	\end{equation*} 
Furthermore, there is no computable sequence of computable numbers $\{u_n\}_{n\in\N}$ with $u_n\geq u_{n+1}$, $n\in\N$ and 
	\begin{equation*}
		\lim_{n\rightarrow\infty}u_n = C(N,\hat{P}).
	\end{equation*}
\end{thm}

\begin{proof}
We consider the non-negative and computable continuous p.s.d. $N$ from Eq. \eqref{eq:n}.

Let $f_1\in(0,\frac{B}{2})$, $f_1\in \R_c$ be arbitrary but fixed. We have that 

	\begin{equation*}
		P_{f_1}=\int_{\frac{B}{2}-f_1}^{\frac{B}{2}+f_1} (N(B+f_1)-N(f))\,df
	\end{equation*}
is the corresponding power concentrated in the interval $[\frac{B}{2}-f_1, \frac{B}{2}+f_1]$. 
It holds that $\hat{P}=P_{f_1}$ and hence $C(N,\hat{P})=C_1(N,f_1)$ however we have already shown that $C_1(N,f_1)\notin \R_c$. This way we have proven the first statement.

Consider the family of recursively enumerable non recursive sets $\{\sA_i\}_{i\in\N}$. This result holds for every $N$ computed from the algorithm for Eq. \eqref{eq:n} that takes as input any recursive function $\varphi_{\sA_i}$ generating a recursively enumerable non-recursive set $\sA_i$. Note that $\hat{P}$ is also a function of $\varphi_{\sA}$, since it depends on $N$, which in turn is determined by $\varphi_{\sA}$.

The proof of the second statement of the theorem follows the same line of argument as in the proof for the second statement of Theorem \ref{thm:cardinalityP}.
\end{proof}

\begin{thm}\label{thm:existenceP}
Let $B>0$, $B\in\R_c$, and $[0,B]$ be arbitrary. There are computable continuous noise spectral densities $N\coloneqq [0,B]\rightarrow \R_{\geq0}^{c}$ such that for every $P>P_*$ and $P\in\R_c$ where
	\begin{equation*}
		P_*=\int_0^B(N(0)-N(f))\, df, 
	\end{equation*}
it always holds that
	\begin{equation*}
		C(N,P)\notin \R_c.
	\end{equation*}
Furthermore, there is no computable sequence of computable numbers $\{u_n\}_{n\in\N}$ with $u_n\geq u_{n+1}$, $n\in\N$ and 
	\begin{equation*}
		\lim_{n\rightarrow\infty}u_n = C(N,P).
	\end{equation*} 
\end{thm}

\begin{proof}
\tco{We consider the non-negative and computable continuous p.s.d. $N$ from Eq. \eqref{eq:n}.}

Let $P>P_*$, $P\in\R_c$ be arbitrary but fixed. 

We have that 
\begin{equation*}
	P = P_*+\Delta B
\end{equation*}

hence $\Delta=\frac{P-P_*}{B}$. Since $P,P_*,B\in\R_c$ then we have that $\Delta\in\R_c$.

We then have that the optimal p.s.d. for $P$ is given by 
\begin{equation*}
	\tcb{P_x^*(f)}=N(0) + \Delta-N(f)
\end{equation*}
for $f\in[0,B]$. This way we have 
\begin{align*}
	C(N,P)&= \int_0^B\ln(\tcb{P_x^*(f)}+N(f))\,df-\int_0^B\ln N(f)\,df\\
	&=B\ln(N(0)+\Delta)-\int_0^B\ln N(f)\,df.
\end{align*}

We have that $B\ln(N(0)+\Delta)\in\R_c$ however we have already shown that $\int_0^B\ln N(f)\,df\notin\R_c$. This implies that $C(N,P)\notin\R_c$, which proves the first statement of the theorem.

Consider the family of recursively enumerable non recursive sets $\{\sA_i\}_{i\in\N}$. This result holds for every $N$ computed from the algorithm for Eq.\eqref{eq:n} that takes any recursive function $\varphi_{\sA_i}$ generating a recursively enumerable non recursive set $\sA_i$. Note that $P_*$ is also a function of $\varphi_{\sA}$, since it depends on $N$, which in turn is determined by $\varphi_{\sA}$. 

To prove the second statement, we have to follow the same line of arguments as in the proof of the second statement of Theorem \ref{thm:capacitycomp}.
\end{proof}

\begin{cor}\label{cor1}
There are infinitely many $P$ with $P\in\R_c$ that fulfill the conditions of Theorem \ref{thm:cardinalityP} or \ref{thm:existenceP}, and for which there is no computable sequence of computable upper-bounds $\{u_n\}_{n\in\N}$ with
	\begin{equation*}
		\lim_{n\rightarrow\infty}u_n = C_1(N,f_1).
	\end{equation*} 
\end{cor} 
\begin{proof}
Assume there is a computable sequence of computable upper bounds $\{u_n\}_{n\in\N}$ with $u_n\geq C(N,P)$ for all $n\in\N$. 
\tco{Consider $N$ from Eq. \eqref{eq:n}.}
Let $\hat{u}_n$ be such that 
	\begin{equation*}
		\hat{u}_n = \min_{1\leq k \leq n}u_k.
	\end{equation*}
$\{\hat{u}_n\}_{n\in\N}$ is a monotonically decreasing computable sequence of computable numbers. It then holds that
 
 	\begin{equation*}
 		\lim_{n\rightarrow \infty}\hat{u}_n = C(N,P).
 	\end{equation*}
\tco{This implies that $C(N,P)$ must be a computable number. However, in the proof of Theorem \ref{thm:capacitycomp}, we have shown that  $C(N,P)\notin \R_c$, leading to the conclusion that our initial assumption must be incorrect.}
\end{proof}
\begin{rem}
Corollary \ref{cor1} states that we can find a computable sequence of achievable rates $\{R_n\}_{n\in\N}$ that converges effectively to the capacity, making the achievability part algorithmically computable. However, it is impossible to algorithmically compute how far the achievable rates $R_n$ are from the capacity. Its implications are beyond the inability to compute a capacity-achieving codebook. Even if we relax the requirement to achieve capacity and allow for some decoding error, it is still impossible to compute an upper bound on the size of the codebook.
\end{rem}
\begin{rem} 
It is interesting to note that while there exist examples of band-limited ACGN channels with computable power spectral densities whose capacities are non-computable numbers, this does not necessarily imply that the converse results of non-computable capacities are also non-computable in general. By non-computable converses, we mean that there is no computable sequence of computable asymptotically sharp upper-bounds. To this end, consider the compound channel. Recent computability studies in \cite{boche2020communication} have shown a converse result: while the compound capacity's converse is computable, i.e., there exist computable sequences of computable upper-bounds that are asymptotically sharp, the achievability of this capacity is not algorithmically computable, i.e., there are no computable sequences of computable lower-bounds that are asymptotically sharp.
\end{rem}

\section{Discussion} 
\label{sec:conclusion}
In this paper, we have focused on studying the algorithmic properties of a simple communication channel: the band-limited ACGN channel. We have shown that there are such channels whose capacities are non-computable numbers. Thus, for a given computable bandwidth, noise power spectrum, and power constraint, there is no algorithm that can effectively compute the capacity of such a channel with a certain desired precision. Moreover, we have also shown that the converse result for those channels is also not algorithmically computable. Specifically, although one can algorithmically construct a sequence of achievable rates that converges to the capacity, it is impossible to compute how far they are from the capacity. So it is impossible to algorithmically compute an upper bound on the size of the codebook for the channel. 

We have also studied the influence of the power constraint on the computability of the capacity of ACGN channels. Unfortunately, we have shown that for those computable channels whose capacity yields a non-computable number adjusting the power constraint does not influence the computability property of the capacity. Moreover, adjusting the power constraint would not enable one to algorithmically compute upper bounds on the capacity.

For more complex channels, such as the FSC, FSC with feedback, and identification of correlation-assisted DMC, it has been shown that the capacity is not Borel-Turing computable, meaning there is no universal algorithm capable of computing the capacity for any channel. However, it is still an open problem whether the capacity of those channels can be computed as a number. By showing that the capacity of this particular ACGN channel is a non-computable number, it immediately implies that the capacity cannot be expressed as a computable function of the channel and power constraint parameters. Therefore, there is no universal algorithm that can take a noise power spectrum, bandwidth, and power constraint as inputs and compute the capacity based on those parameters.

As future work, one could aim to determine the set of channels for which the capacity yields a computable number. In such cases, it could be possible to express the capacity as a computable function of the channel and power constraint parameters, and then study its computational complexity. This would allow for the development of algorithms to compute the capacity efficiently.

\balance
\bibliographystyle{IEEEtran}
\bibliography{IEEEabrv,confs-jrnls,references_coloregaussian}

\begin{thebibliography}{10}
\providecommand{\url}[1]{#1}
\csname url@samestyle\endcsname
\providecommand{\newblock}{\relax}
\providecommand{\bibinfo}[2]{#2}
\providecommand{\BIBentrySTDinterwordspacing}{\spaceskip=0pt\relax}
\providecommand{\BIBentryALTinterwordstretchfactor}{4}
\providecommand{\BIBentryALTinterwordspacing}{\spaceskip=\fontdimen2\font plus
\BIBentryALTinterwordstretchfactor\fontdimen3\font minus
  \fontdimen4\font\relax}
\providecommand{\BIBforeignlanguage}[2]{{%
\expandafter\ifx\csname l@#1\endcsname\relax
\typeout{** WARNING: IEEEtran.bst: No hyphenation pattern has been}%
\typeout{** loaded for the language `#1'. Using the pattern for}%
\typeout{** the default language instead.}%
\else
\language=\csname l@#1\endcsname
\fi
#2}}
\providecommand{\BIBdecl}{\relax}
\BIBdecl

\bibitem{fettweis20216gtactile}
G.~P. Fettweis and H.~Boche, ``6{G}: The personal {T}actile {I}nternet—and
  open questions for information theory,'' \emph{BITS Inf. Theory Mag.},
  vol.~1, no.~1, pp. 71--82, 2021.

\bibitem{fettweis20226gtrust}
------, ``On 6{G} and trustworthiness,'' \emph{Commun. ACM}, vol.~65, no.~4,
  pp. 48--49, 2022.

\bibitem{shannon1967lower}
C.~E. Shannon, R.~G. Gallager, and E.~R. Berlekamp, ``Lower bounds to error
  probability for coding on discrete memoryless channels. {I},'' \emph{Inf.
  Contr.}, vol.~10, no.~1, pp. 65--103, 1967.

\bibitem{arimoto1972algorithm}
S.~Arimoto, ``An algorithm for computing the capacity of arbitrary discrete
  memoryless channels,'' \emph{{IEEE} Trans. Inf. Theory}, vol.~18, no.~1, pp.
  14--20, Jan. 1972.

\bibitem{blahut1972computation}
R.~Blahut, ``Computation of channel capacity and rate-distortion functions,''
  \emph{{IEEE} Trans. Inf. Theory}, vol.~18, no.~4, pp. 460--473, Jul. 1972.

\bibitem{boche2022algorithmic}
H.~Boche, R.~F. Schaefer, and H.~V. Poor, ``Algorithmic computability and
  approximability of capacity-achieving input distributions,'' (in press).

\bibitem{boche2020shannon}
------, ``Shannon meets {T}uring: Non-computability and non-approximability of
  the finite state channel capacity,'' \emph{Commun. Inf. Syst.}, vol.~20,
  no.~2, pp. 81--116, 2020.

\bibitem{grigorescu2022capacity}
A.~Grigorescu, H.~Boche, R.~F. Schaefer, and H.~V. Poor, ``Capacity of finite
  state channels with feedback: Algorithmic and optimization theoretic
  properties,'' \emph{arXiv preprint arXiv:2201.11639}, 2022.

\bibitem{boche2020communication}
H.~Boche, R.~F. Schaefer, and H.~V. Poor, ``Communication under channel
  uncertainty: An algorithmic perspective and effective construction,''
  \emph{{IEEE} Trans. Signal Process.}, vol.~68, pp. 6224--6239, 2020.

\bibitem{boche2019identification}
------, ``Identification capacity of correlation-assisted discrete memoryless
  channels: Analytical properties and representations,'' in \emph{Proc. IEEE
  Int. Symp. Inf. Theory}, Paris, France, Jul. 2019, pp. 470--474.

\bibitem{shannon1949communication}
C.~E. Shannon, ``Communication theory of secrecy systems,'' \emph{Bell Syst.
  Tech.~J.}, vol.~28, no.~4, pp. 656--715, Oct. 1949.

\bibitem{shannon1949communicationnoise}
------, ``Communication in the presence of noise,'' \emph{Proc. IRE}, vol.~37,
  no.~1, pp. 10--21, 1949.

\bibitem{wyner1966capacity}
A.~D. Wyner, ``The capacity of the band-limited gaussian channel,'' \emph{Bell
  Syst. Tech.~J.}, vol.~45, no.~3, pp. 359--395, 1966.

\bibitem{ash1963capacity}
R.~B. Ash, ``Capacity and error bounds for a time-continuous {G}aussian
  channel,'' \emph{Inf. Contr.}, vol.~6, no.~1, pp. 14--27, 1963.

\bibitem{shannon1959probability}
C.~E. Shannon, ``Probability of error for optimal codes in a {G}aussian
  channel,'' \emph{Bell Syst. Tech.~J.}, vol.~38, no.~3, pp. 611--656, 1959.

\bibitem{ash2012information}
R.~B. Ash, \emph{Information {T}heory}.\hskip 1em plus 0.5em minus 0.4em\relax
  Courier Corporation, 2012.

\bibitem{cover1999elements}
T.~M. Cover, \emph{Elements of {I}nformation {T}heory}.\hskip 1em plus 0.5em
  minus 0.4em\relax John Wiley \& Sons, 1999.

\bibitem{ihara1993information}
S.~Ihara, \emph{Information {T}heory for {C}ontinuous {S}ystems}.\hskip 1em
  plus 0.5em minus 0.4em\relax World Scientific, 1993, vol.~2.

\bibitem{tse2005fundamentals}
D.~Tse and P.~Viswanath, \emph{Fundamentals of {W}ireless
  {C}ommunication}.\hskip 1em plus 0.5em minus 0.4em\relax Cambridge university
  press, 2005.

\bibitem{pinsker1964information}
M.~S. Pinsker, \emph{Information and {I}nformation {S}tability of {R}andom
  {V}ariables and {P}rocesses}.\hskip 1em plus 0.5em minus 0.4em\relax
  Holden-Day, 1964.

\bibitem{gallager1968information}
R.~G. Gallager, \emph{Information Theory and Reliable Communication}.\hskip 1em
  plus 0.5em minus 0.4em\relax John Wiley \& Sons, Inc., 1968.

\bibitem{forney1998modulation}
G.~Forney and G.~Ungerboeck, ``Modulation and coding for linear {G}aussian
  channels,'' \emph{{IEEE} Trans. Inf. Theory}, vol.~44, no.~6, pp. 2384--2415,
  Oct. 1998.

\bibitem{goldsmith2001capacity}
A.~J. Goldsmith and M.~Effros, ``The capacity region of broadcast channels with
  intersymbol interference and colored {G}aussian noise,'' \emph{{IEEE} Trans.
  Inf. Theory}, vol.~47, no.~1, pp. 219--240, Jan. 2001.

\bibitem{hughes1975capacity}
D.~Hughes-Hartogs, ``The capacity of the {D}egraded {S}pectral {G}aussian
  {B}roadcast {C}hannel.'' Ph.D. dissertation, 1975.

\bibitem{poltyrev1979capacity}
G.~S. Poltyrev, ``Capacity for a sum of broadcast channels,'' \emph{Problemy
  Peredachi Informatsii}, vol.~15, no.~2, pp. 40--44, 1979.

\bibitem{el1980capacity}
A.~El~Gamal, ``Capacity of the product and sum of two unmatched broadcast
  channels,'' \emph{Problemy Peredachi Informatsii}, vol.~16, no.~1, pp. 3--23,
  1980.

\bibitem{hirt1988capacity}
W.~Hirt and J.~L. Massey, ``Capacity of the discrete-time {G}aussian channel
  with intersymbol interference,'' \emph{{IEEE} Trans. Inf. Theory}, vol.~34,
  no.~3, pp. 38--38, 1988.

\bibitem{verdu1989multiple}
S.~Verdu, ``Multiple-access channels with memory with and without frame
  synchronism,'' \emph{{IEEE} Trans. Inf. Theory}, vol.~35, no.~3, pp.
  605--619, 1989.

\bibitem{cheng1993gaussian}
R.~S. Cheng and S.~Verd{\'u}, ``Gaussian multiaccess channels with {ISI}:
  Capacity region and multiuser water-filling,'' \emph{{IEEE} Trans. Inf.
  Theory}, vol.~39, no.~3, pp. 773--785, May 1993.

\bibitem{turing1936computable}
A.~M. Turing \emph{et~al.}, ``On computable numbers, with an application to the
  {E}ntscheidungsproblem,'' \emph{Proc. London Math. Soc.}, vol.~2, no.~42, pp.
  230--265, 1936.

\bibitem{turing1938computable}
A.~M. Turing, ``On computable numbers, with an application to the
  {E}ntscheidungsproblem. a correction,'' \emph{Proc. London Math. Soc.},
  vol.~2, no.~43, pp. 544--546, 1937.

\bibitem{weihrauch2000computable}
K.~Weihrauch, \emph{Computable {A}nalysis: {A}n {I}ntroduction}.\hskip 1em plus
  0.5em minus 0.4em\relax Springer Science \& Business Media, 2000.

\bibitem{godel1930vollstandigkeit}
K.~G{\"o}del, ``Die {V}ollst{\"a}ndigkeit der {A}xiome des logischen
  {F}unktionenkalk{\"u}ls,'' \emph{Monatshefte f{\"u}r Mathematik und Physik},
  vol.~37, no.~1, pp. 349--360, 1930.

\bibitem{godel1934undecidable}
------, ``On undecidable propositions of formal mathematical systems,
  mimeographed lecture notes by {S}tephen {C}. {K}leene and {J}. {B}arkley
  rosser,'' pp. 39--74, 1934.

\bibitem{kleene1952introduction}
S.~C. Kleene, \emph{Introduction to {M}etamathematics}.\hskip 1em plus 0.5em
  minus 0.4em\relax Amsterdam : North-Holland Publishing ; Groningen : P.
  Noordhoff N.V., 1952.

\bibitem{minsky1961recursive}
M.~L. Minsky, ``Recursive unsolvability of {P}ost's problem of "tag" and other
  topics in theory of {T}uring machines,'' \emph{Annals Math.}, pp. 437--455,
  1961.

\bibitem{avigad2014computability}
J.~Avigad, V.~Brattka, and R.~Downey, \emph{Computability and {A}nalysis: {T}he
  {L}egacy of {A}lan {T}uring.}\hskip 1em plus 0.5em minus 0.4em\relax
  Cambridge, UK: Cambridge University Press, 2014.

\bibitem{proakis2001digital}
J.~G. Proakis and M.~Salehi, \emph{Digital {C}ommunications}.\hskip 1em plus
  0.5em minus 0.4em\relax McGraw-hill New York, 2001, vol.~4.

\bibitem{haykin2001communication}
S.~Haykin, \emph{Communication {S}ystems /}, 4th~ed.\hskip 1em plus 0.5em minus
  0.4em\relax John Wiley \& Sons.,, 2001.

\bibitem{flanagan2006proving}
M.~F. Flanagan, ``On proving the water pouring theorem for information rate
  optimization,'' in \emph{International Conference on Signals and Electronic
  Systems}.\hskip 1em plus 0.5em minus 0.4em\relax Citeseer, 2006.

\bibitem{pour2017computability}
M.~B. Pour-El and J.~I. Richards, \emph{Computability in {A}nalysis and
  {P}hysics}.\hskip 1em plus 0.5em minus 0.4em\relax Cambridge University
  Press, 2017.

\bibitem{soare1978recursively}
R.~I. Soare, ``Recursively enumerable sets and degrees,'' \emph{Bulletin of the
  American Mathematical Society}, vol.~84, no.~6, pp. 1149--1181, 1978.

\end{thebibliography}
\end{document}